\documentclass[showpacs,twocolumn,prb]{revtex4-1}
\usepackage{graphicx}
\usepackage{subfloat}

\newcommand{\be}{\begin{equation}}
\newcommand{\ee}{\end{equation}}
\newcommand{\bea}{\begin{eqnarray}}
\newcommand{\eea}{\end{eqnarray}}
\newcommand{\bwt}{\begin{widetext}}
\newcommand{\ewt}{\end{widetext}}

\usepackage{color}

\newcommand{\bi}{\bf i}
\newcommand{\bj}{\bf j}

\begin{document}

\title{BLF-SSH Polarons coupled to Acoustic Phonons in the Adiabatic limit}

\author{Carl J. Chandler and F. Marsiglio}
\affiliation{Department of Physics, University of Alberta, Edmonton, Alberta, Canada, T6G~2G7}

\begin{abstract}
We survey polaron formation in the BLF-SSH model using acoustic phonons in the adiabatic limit.
Multiple different numerical optimization routines and strong coupling analytical calculations
are used to find a robust ground state energy
for a wide range of coupling strengths. The electronic configuration
and accompanying ionic distortions of the polaron were determined, as well 
as a non-zero critical coupling strength for polaron formation in two and three dimensions. 
\end{abstract}

\pacs{}
\date{\today }
\maketitle

\section{Introduction}
Many-body calculations of electrons coupled to acoustic phonon modes were first proposed by Barisi\'{c}, Labb\'{e}, and Friedel (BLF)\cite{barisic70} in the context of understanding transition metal superconductivity in 1970. The same coupling was
subsequently reintroduced by Su, Schrieffer, and Heeger (SSH)\cite{su79,su80} 10 years later to model soliton modes in long 
polyacetylene chains. More recently there has been a revival of interest in these types of models to describe 
superconductivity in the cuprate materials, though typically only the so-called BLF-SSH form of the coupling is adopted. For
both physical and technical reasons, the acoustic phonons are usually modelled as Einstein oscillators, i.e. optical modes.\cite{csgmodel} The BLF-SSH model has also been used recently in problems concerning conducting polymers for electronic and solar-cell applications,\cite{solar} as well as problems in biophysics.\cite{biophysics}

The BLF-SSH model differs from the commonly used Holstein model\cite{holstein59} in two main ways. 
First, as already mentioned, it uses acoustic phonon modes, thus maintaining relevance for materials 
without optical modes. Second, the electron-phonon interaction in the BLF-SSH model 
modifies the electron hopping term, not the on-site energy as in the Holstein model. Both these modifications makes the
BLF-SSH model technically more difficult, but they also potentially alter the physics somewhat, as the lowering
of energy due to the electron-phonon coupling is associated with {\em movement} of the electron, and not with
the (Coulombic) potential energy between the electron and the displaced ionic charge. 
The so-called CSG model\cite{csgmodel} shows somewhat unusual properties, even for the single polaron, presumably
due to the optical mode simplification.

Much of the work done on this model is in the one dimensional adiabatic approximation, i.e. the phonons are treated classically.\cite{su79,su80} 
Barisi\'{c}, Labb\'{e}, and Friedel\cite{barisic70} used BCS and diagrammatic methods to address superconductivity, but after the SSH revival, the
effects of quantum fluctuations were examined through quantum Monte Carlo and 
renormalization group studies,\cite{hirsch82} and these authors focused on half-filling.  
They found that the lattice ordering (in one dimension) was reduced by quantum fluctuations.

Further studies were performed for a single polaron, based on variational calculations,\cite{peeters85} for the Fr\"ohlich
Hamiltonian in the continuum with acoustic phonons and with a wave vector cutoff to mimic lattice effects. These authors generally found a phase transition to a ``self-trapping'' state, as a function of coupling strength.

For the BLF-SSH model, however, very little work has been done in the quantum regime for a single electron. We have studied the BLF-SSH polaron using perturbation theory, and were unable to find, for example, a perturbative regime in one dimension where polaron effects are absent.\cite{li11} In Ref.~\onlinecite{miyasaka01} the properties of a single polaron in the BLF-SSH model 
have been studied in one and two dimensions, using the adiabatic approximation. 
Unfortunately, we believe this 2D study has serious errors, and their results display
unphysical emergent phenomena (see below). Here we will present a comprehensive survey 
of the  adiabatic BLF-SSH model in one, two and three dimensions. 

In the adiabatic limit the electrons
are treated quantum mechanically, while the ions are treated semi-classically. The ions are considered to have 
no kinetic energy and their displacements from equilibrium are treated as input parameters to the Hamiltonian.
Since the electronic bandwidth in real materials is often very large compared to the phonon energy
scale, this limit is expected to be physically relevant. On the other hand certain pathologies have come to be associated
with the adiabatic limit. For example, as will be reported below, we found a critical coupling strength in dimensions higher than one, beyond which
the electron forms a polaron-like ground state, and below which the electron is decoupled from the lattice. From studies
of the Holstein model,\cite{kabanov93,bonca99,li10}, the existence of a critical coupling strength is expected to not survive away from the adiabatic limit.
Nonetheless, studies of the adiabatic limit give a good picture of what will occur in the near-adiabatic limit, particularly in the
strong coupling limit.

This paper is organized as follows. In the next section we define the model and the adiabatic approximation, and follow this
with a short discussion concerning our methods. We then display some analytical results, and follow up with numerical
results in the ensuing section. In the final section we provide a brief summary.

\section{The Model}

We begin by writing down the Hamiltonian for a 2D system - this is readily generalized to
the 1D and 3D cases that are also treated in this paper: 

\begin{eqnarray}
H &=& -\sum_{\langle \bi,\bj \rangle} \ t_{\bi \bj } \biggl(c_{\bi\sigma}^\dagger c_{\bj\sigma} + h.c. \biggr) + \sum_{\bi} \biggl[ {p_{x\bi}^2 \over 2M} + {p_{y\bi}^2 \over 2M} \biggr]\nonumber \\
&+&  {1 \over 2} K\sum_{\langle \bi,\bj \rangle} \biggl[ \bigl(u_{x\bi} - u_{x\bj}\bigr)^2 +  \bigl(u_{y\bi} - u_{y\bj}\bigr)^2  \biggr],
\label{ham_xy}
\end{eqnarray}
where angular brackets denote nearest neighbours only without double counting, and the $\bi$ and $\bj$ indices are written in boldface to emphasize that for the
$D$-dimensional case they are $D$-dimensional vectors. 
The operators and parameters are as follows: $c_{\bi\sigma}^\dagger$ ($c_{\bi\sigma}$) creates (annihilates) an electron at site $\bi$ with spin $\sigma$. The $x$-components for the ion momentum and displacement are given by $p_{x\bi}$, and displacement $u_{x\bi}$, respectively (similarly for the $y$-components), and the ions have mass $M$ and spring constant $K$ connecting nearest neighbours only. Furthermore, 
\begin{equation}
t_{\bi\bj}  = t -  \alpha (u_{x\bi} - u_{x\bj})\delta_{\bi,\bj\pm \hat{a}_x} - \alpha (u_{y\bi} - u_{y\bj})\delta_{\bi,\bj\pm \hat{a}_y}.
\label{hopping}
\end{equation}
Note that the parameter $\alpha$ can be written as a derivative of the hopping amplitude with respect to displacement.
Here it is simply treated as a parameter. Moreover, the electron hopping is modified only by ionic motions in the same
direction, i.e. longitudinal coupling only, consistent with an expansion of the coupling term to linear order only in the displacements.\cite{berciu13}
The adiabatic approximation is achieved by dropping the kinetic energy term for the ions: 
\begin{eqnarray}
H &=& -\sum_{\langle \bi,\bj \rangle} \ t_{\bi\bj} \biggl(c_{\bi\sigma}^\dagger c_{\bj\sigma} + h.c. \biggr) \nonumber \\
&+&  {1 \over 2} K\sum_{\langle \bi,\bj \rangle} \biggl[ \bigl(u_{x\bi} - u_{x\bj}\bigr)^2 +  \bigl(u_{y\bi} - u_{y\bj}\bigr)^2  \biggr].
\label{ham_adiabatic_xy}
\end{eqnarray}
This means that we can treat the ionic displacements as $c$-numbers, and the electronic part of the Hamiltonian remains as an
eigenvalue problem. We change variables for the ions, since the Hamiltonian depends only on the 
separation between the ion sites. Thus we define $\tilde{x}_{\bi} = u_{x\bj} - u_{\bi+\bf{\delta}_x}$ and  $\tilde{y}_{\bi} = u_{y\bi} - u_{y\bi+{\bf{\delta}_y}}$. 

\begin{figure}
\begin{center}
\includegraphics[height=0.8in,width=2.1in]{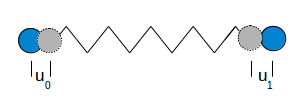}
\caption{ A one dimensional depiction of the variables used to describe ionic motion. 
The full blue circles are ions at their equilibrium positions, and the grey dotted circles are ions displaced from their 
equilibrium positions. The $u_{x\bi}$ variables are then seen to be the displacements from equilibrium and $\tilde{x}_{\bf{0}} = u_{x\bf{0}} - u_{x\bf{1}}$ so $\tilde{x}_{x\bf{0}}$ can be thought of as the distance change between the two ions after subtracting the 
equilibrium distance spacing.}
\label{fig:sshDef}
\end{center}
\end{figure}

This simplifies the calculations and somewhat changes the nature of the boundary conditions. We will use
periodic boundary conditions; in the original $u_{x\bi}$ and $u_{y\bi}$ variables, this would mean that if a disruption occurred somewhere in the lattice
(say, near the electron), then this `disruption' would have to `heal' itself at the boundary. By switching to the $\tilde{x}$ and $\tilde{y}$
variables this is no longer true. A separation of ions near the electron would simply `push' the remaining ions further out. We have effectively
eliminated the mode that corresponds to uniform translation of all the ions and introduced a stretching mode that allows the entire lattice to expand or contract
(this is not possible with conventional periodic boundary conditions).
In the thermodynamic limit this choice of variable and boundary conditions does not effect the physical result (an electron distorting the ions in its
vicinity), but use of the $\tilde{x}_{\bi}$ and $\tilde{y}_{\bi}$ variables results greatly reduces finite size effects for systems smaller than the thermodynamic limit.

It is advantageous to rescale the ion displacement parameters as dimensionless variables. To this end we define
\begin{equation}
 x_{\bi} = \frac{\alpha}{K} \tilde{x}_{\bi}
\end{equation}
\begin{equation}
 y_{\bi} = \frac{\alpha}{K}  \tilde{y}_{\bi}
\end{equation}
As is customary we define a dimensionless electron phonon coupling strength $\lambda$: 
\begin{equation}
 \lambda = \frac{\alpha^2}{\omega_0^2 MW},
\end{equation}
where $W \equiv 4Dt$ is the electronic bandwidth for a `cubic' lattice in $D$-dimensions, and 
\begin{equation}
\omega_0 = \sqrt{ \frac{4K}{M} }.
\end{equation}
Thus the adiabatic Hamiltonian becomes: 
\begin{eqnarray}
 H &=& \sum_{\bi}  [-t + \lambda W x_{\bi}] [c_{\bi}^\dagger c_{\bi + {\bf{\delta}_x}} + h.c. ] \nonumber \\
  &+& \sum_{\bi}  [-t + \lambda W y_{\bi}][ c_{\bi}^\dagger c_{\bi + {\bf{\delta}_y}} + h.c. ] \nonumber \\
 &+&  \sum_{\bi} {\lambda W \over 2} ({x}_{\bi}^2 +  {y}_{\bi}^2 ).
 \label{ham_adiab}
\end{eqnarray}

Consideration of only longitudinal modes keeps the directions independent of one other, 
consistent with what is generally done in the fully quantum mechanical treatment.\cite{li11}
This also neglects changes in $y$-distances
that could come from changing nearby $x$-distances through triangulation. This is 
justifiable since the change in the hopping due to the electron-phonon interaction 
is itself a linear approximation and these definitions are consistently linear 
in the changes in bond length.

The adiabatic BLF-SSH model (with coupling to longitudinal modes as described here)
has been studied in one dimension\cite{su79} and two dimensions.\cite{miyasaka01} More recent studies
are motivated by biophysical and polymer applications and are generally done in one dimension.\cite{Conwell00,biophysics,solar}
Calculations in the adiabatic limit are useful since they allow us to understand the physical structure of a 
polaron, both electronically, and through the accompanying ionic displacements. 

\section{Methods}

With the Hamiltonian defined in Eq.~(\ref{ham_adiab}), for a given electronic wave function the ground state energy can be determined
and minimized with respect to the ionic displacements. Solving the semi-classical adiabatic model is therefore a problem of function minimization. 
In this formulation, the bond length parameters are the variational input parameters. Given their values, the electron energy can be evaluated by 
evaluating the tight binding electronic Hamiltonian matrix elements with the $t_{\bi}$ values calculated from the bond lengths. 
The ionic energy is a simple classical sum of the bond lengths squared and together these terms give the binding energy of
the polaron. The number of parameters then scales as the number of ion sites and also dimension. 

The multivariable minimization problem is in general much easier than the full quantum many body problem, but remains a 
very difficult problem in its own right. Finding a solution with a low energy is not particularly difficult, 
but knowing that one has the lowest possible solution is virtually impossible. This is at the root of the 
confusion in the field --- there is no good way to distinguish whether one has found the global minimum, or simply a low local minimum. 
For many applications, such as the traveling salesman problem, finding a solution that is quite close to the absolute 
minimum is acceptable. However, for the polaron, two solutions with similar energies may have a very different physical 
structure; thus finding the global minimum is important for a proper physical understanding. 

There are many different algorithms for multivariable minimization and to establish confidence in our results
we have implemented several different ones, verifying that we have the correct answer. Each has its own 
strengths and weaknesses and by combining them we have a much better understanding of the energy landscape. 

The state of the art in multivariable minimization with no a priori knowledge is the genetic algorithm. \cite{DEbook}
This algorithm is very effective at searching through the entire space for low energy solutions, but is rather slow
for large numbers of parameters and has a hard time ``fine tuning'' a solution. The basic method is to create 
a population of points in the N-dimensional space, then allow them to breed, where they swap coordinates, and
small random variations are introduced. These new points are then ranked by evaluating their energies, and the best
half are allowed to breed and compose the next iteration's population. 

Alternatively, one may use minimization algorithms using the gradient. 
By using the Hellman-Feynman theorem, the gradient can be computed as a function of the bond lengths and the eigenvector 
of the tight-binding Hamiltonian can be constructed by using those bond lengths. We have 
\begin{equation}
 \frac{\partial E }{\partial {y_{\bi}}} = \langle \psi | \frac{ \partial H_{elec} }{\partial {y_{\bi}}} | \psi \rangle + \lambda W {y_{\bi}}.
\end{equation}
and similarly for the $x_{\bi}$ parameters.
There are two ways to use this information. First, 
one can set up a self-consistent set of equations and iterate through these. Second, one can use a conjugate gradient
optimization routine. Both of these methods run much faster than the genetic algorithm, but they do not sample
more than one point in configuration space at a time. This makes them more prone to falling into local minima. 

We implemented all of these algorithms and found the best performance from the conjugate gradient method. It found the
same configuration as the genetic algorithm on small systems given random
initial conditions, and could handle larger systems with ease. It was, however, more sensitive to errors 
in the eigenvalue and eigenvector of the diagonalization routine than the iterative method.

Preliminary calculations showed that the polaron would be very small at strong coupling, so we first performed searches of 
the solution for strong coupling parameters on small clusters.  We used both the differential algorithm and conjugate gradient
algorithm using random starting conditions. This generally produced a few low energy configurations. We then used these configurations
as starting conditions for a sweep towards zero coupling strength. We would find the lowest energy configuration
for a given $\lambda$, and then use that configuration as the starting point for the next lower  $\lambda$ calculation. 
This avoided getting lost in multi-dimensional phase space. We did other surveys for starting configurations that were not 
low energy solutions at strong coupling, but in these cases only the trivial solution of a free electron immersed in a lattice of unstretched
bonds was found. 


\section{ Analytical Results} 

Before we present data from our numerical simulations, it is instructive to examine 
analytically the case of strong coupling. Here the polaron is very small, and thus
we can perform a simple analytical calculation to obtain the optimum solution.

In one dimension, instead of following the numerical procedure of periodic boundary conditions,
we adopt open boundary conditions, since we are anticipating a very small polaron.
There is a general distinction between chains with an even or odd number of sites.
For example, in the two-site model the electron wave function is expected to be a symmetric
linear combination of the electron on both sites, i.e. $|\psi_2\rangle = (c_0^\dagger |0\rangle + c_1^\dagger |0 \rangle)/\sqrt{2}$. 
The subscript 0 (1) refers to the left (right) site.
The problem is immediately diagonal, and the electronic energy from the electronic Hamiltonian is $\epsilon_{\rm el} = -(t - \lambda W x)$, where
$x$ represents the dimensionless `stretch' of the one bond in the problem. Combined with the ionic part of the Hamiltonian we
obtain a total energy of $E = -t + \lambda W x + \lambda W x^2/2$.  Minimization gives $x = -1$ and $E_{\rm min} = -t - \lambda W/2$.

For three sites there are two independent normalized wave functions, 
\bea
|\phi_0 \rangle & = & c_0^\dagger|0\rangle \nonumber \\
|\phi_1 \rangle & = & {1 \over \sqrt{2}}\bigl( c_{-1}^\dagger + c_1^\dagger \bigr) | 0\rangle,
\label{basis}
\eea
where $-1$, $0$ and $1$ represent the site indices,
and the electron wave function $|\psi_3\rangle$ is given in terms of these two basis states:
\be
|\psi_3\rangle = a_0 |\phi_0\rangle + a_1 | \phi_1 \rangle
\label{psi3}
\ee
We therefore have an eigenvalue problem for the two coefficients and the electronic energy, $\epsilon_{\rm el}$, as
   \begin{equation}
   \left (
   \begin{array}{cc}
   0 & -\sqrt{2} (t - \lambda W x_0) \\
   -\sqrt{2}  (t - \lambda W x_0) & 0
   \end{array}
   \right )
   \left (
   \begin{array}{cc}
   a_0 \\
   a_1
   \end{array} 
   \right )
   = \epsilon
   \left (
   \begin{array}{cc}
   a_0 \\
   a_1
   \end{array} 
   \right ),
   \end{equation}
where, due to the symmetry of the problem, the dimensionless bond stretches on the left and on the right will be equal (denoted here by $x_0$).  The eigenvalues are readily determined, with the electronic ground state energy given by
$\epsilon_{\rm el} = -\sqrt{2} (t - \lambda x_0 )$; then the total energy
\be
E_{\rm tot} = -\sqrt{2} (t - \lambda x_0 ) + \lambda W x_0^2
\label{tot_energy3}
\ee
is minimized by $x_0 = -1/\sqrt{2}$; this gives
\be
E_{\rm GS} = - \sqrt{2}t - {1 \over 2}\lambda W.
\label{ee_3}
\ee
This represents a lower energy than the 2 site model, and in general the solution with an odd number of sites partially occupied by the
electron will have a lower energy than that with an even number of sites. The eigenvector corresponding to this energy
is 
 \be
|\phi_{\rm GS} \rangle  =  {1 \over {2}} c_{-1}^\dagger|0\rangle +  {1 \over \sqrt{2}} c_0^\dagger | 0\rangle + {1 \over {2}} c_{1}^\dagger |0\rangle,
\label{gseig}
\ee
and corresponds to a central maximum electron amplitude with two smaller amplitudes on either side. One relative ion displacement is
required, with $x_0 = -1/\sqrt{2}$. This corresponds to the ions on either side of the central maximum moving closer to the centre, with all
other ions on either side following suit, meaning that there are no further relative displacements. For reasons further explained in the Appendix,
we expect this solution to properly represent the strong coupling solution, even for the quantum case. To verify this for the adiabatic limit, at least,
we expand the Hilbert space.

\begin{figure}
\begin{center}
\includegraphics[height=2.8in,width=3.7in]{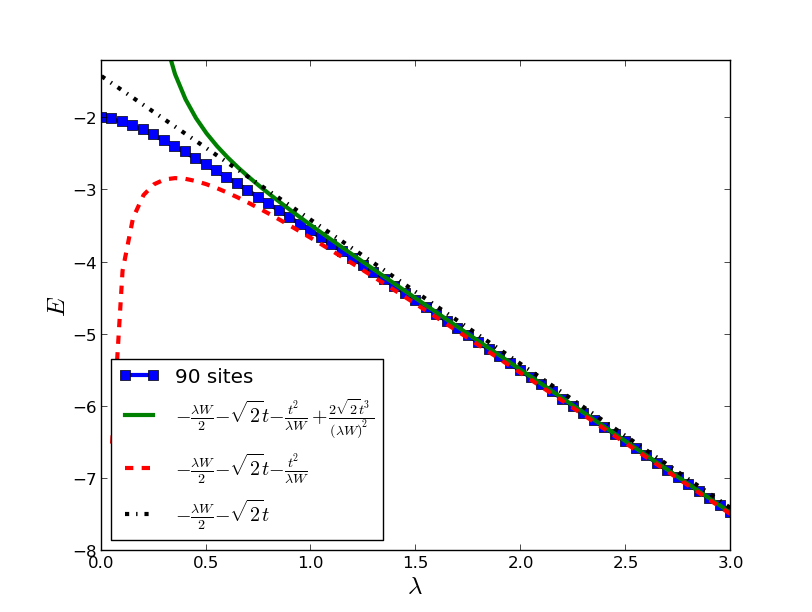}
\caption{The ground state energy for the 1D BLF-SSH model in the adiabatic limit. The solid curve with (blue) squares
is the numerical solution, while the various strong coupling expansions are shown as indicated in the legend. The 
divergence in weak coupling is due to the finite expansion in $t/\lambda W$ and not from some effect of the model itself.
}
\label{fig:1Danalytical}
\end{center}
\end{figure}

With five sites, we use an additional electron basis state,  
\be
|\phi_2 \rangle  =  {1 \over \sqrt{2}}\bigl( c_{-2}^\dagger + c_2^\dagger \bigr) | 0\rangle,
\label{basis2}
\ee
and an additional stretch denoted by $x_1$, between site 1 and site 2 (or site -1 and site -2).
The equations are slightly more complicated, as a $3 \times 3$ matrix must be diagonalized,
   \begin{equation}
   \left (
 \begin{array}{ccc}
   0 & -\sqrt{2}t_0 & 0\\
   -\sqrt{2} t_0 & 0 & -t_1 \\
   0 & -t_1 & 0
   \end{array}
   \right )
   \left (
   \begin{array}{cc}
   a_0 \\
   a_1 \\
   a_2 \\
   \end{array} 
   \right )
   = \epsilon
   \left (
   \begin{array}{cc}
   a_0 \\
   a_1\\
   a_2
   \end{array} 
   \right ),
   \end{equation}
where $a_0$ and $a_1$ are amplitudes of the two basis states in Eq.~(\ref{basis}) as before and
$a_2$ is the amplitude for the basis state $|\phi_2\rangle$, and $t_k \equiv t-\lambda W x_k$ for $k=0,1$.
A straightforward diagonalization gives 
\be
\epsilon_{\rm el} = -\sqrt{2(t-\lambda W x_0)^2 + (t - \lambda W x_1)^2},
\label{ele_ener3}
\ee
so that the total energy is well-defined in terms of $x_0$ and $x_1$. Taking partial derivatives with respect
to these two parameters and setting them to zero then gives two equations that cannot be solved in closed form.
However, an expansion in increasing powers of $t/(\lambda W)$ gives a minimum energy
\begin{equation}
 E = \frac{-\lambda W}{2} - \sqrt{2}t - \frac{t^2}{\lambda W} + \frac{2 \sqrt{2} t^3}{(\lambda W)^2} +  \ O(\frac{t}{\lambda W} )^3 .
\end{equation}
Note that the corresponding eigenvector is given by Eq.~(\ref{gseig}), {\it plus} corrections of order $O(1/\lambda)$, including the
amplitude on the two sites furthest from the centre. Furthermore $x_1 = O (1/\lambda)$, and $x_0 = -1/\sqrt{2} + O (1/\lambda)$.
This confirms Eqs.~(\ref{gseig}) and (\ref{ee_3}) as the strong coupling solutions.
Figure ~\ref{fig:1Danalytical}  shows these solutions along with our numerical solution for the thermodynamic limit, and the agreement is
very good for strong coupling.

Since all the corrections beyond the three-site model are of order $t/\lambda W$ or higher we are assured 
that the three-site solution represents the true strong coupling limit. There are no finite size effects 
that need to be included and the minimization is analytical and thus this solution is not subject to being trapped in 
a local minimum as numerical methods can be. 

Using the same analytical methods we can find the strong coupling limit for the 2D and 3D cases as well.
We restrict ourselves to solutions that have the same symmetries as the lattice since a symmetric 
Hamiltonian is expected to have symmetric solutions. Nonetheless, there are alternative analogues to the two and three 
site configurations that we considered in one dimension. The two general types of configurations are those in which a
polaron is centred on a site, or those in which a polaron is centred at a location which is not a site; instead it is centred on a point 
equidistant from all neighbouring sites. In one dimension these configurations were the cases with an odd or even number of
sites with significant amplitude, respectively. Some details along the lines given above for the one dimensional case are
given in the Appendix. 
Here we simply present 
the final energies and configurations.  The calculations are straightforward; numerical results are obtained by diagonalization of
finite systems until convergence is achieved, and analytical results  can be done very quickly with the help of a 
computer algebra system like Mathematica. 

In two dimensions a curious degeneracy occurs --- the two potential configurations ( Fig. ~\ref{fig:2Dconfigs}) have the same energy in the extreme strong
coupling limit (see Appendix),
\begin{equation}
 E = \frac{-\lambda W }{2} - 2t 
\end{equation}

\begin{figure}
\begin{center}
\includegraphics[height=2.0in,width=3.5in]{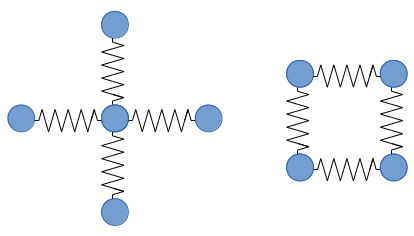}
\caption{ The two possible configurations for stretched bonds in the 2D strong coupling limit. The square has the electron amplitude
equal on all four sites while the star configuration has half of the electron probability on the centre site and one eighth on each of the 
four surrounding sites. }
\label{fig:2Dconfigs}
\end{center}
\end{figure}

However, it should be noted that this degeneracy is really a special case applicable to the simple square lattice. For example the 2D 
honeycomb lattice also has two analogous solutions, but they are not degenerate (see Appendix).
For the two dimensional square lattice this degeneracy remains for all coupling strengths, as our numerical results showed no discernible
difference once converged for finite size effects. This appears to come from the fact that both solutions have the 
same number of bonds stretched on the square lattice so the energy cost is the same, and they have the same electron 
energy even though they have different electron configurations in real space. The existence of two solutions may 
be more important in many-electron calculations since they have different physical sizes, but for single 
electron studies they seem to be interchangeable. 

In 3D there are again two types of solutions, those centred on an actual site of the lattice (so-called `star' configuration) and and those centred on
a point which would be the centre of a cube of eight sites. We found the 'star' configuration to always have the lowest energy; in strong coupling the two energies are given by Eqs.~(\ref{ee_3d_app}):
\begin{eqnarray}
E_{\rm star}& = &-{1 \over 2}\lambda W - \sqrt{6}t  \phantom{aaaa} {\rm (3D)} \nonumber \\
E_{\rm cube}& = &-{3 \over 8}\lambda W - 3t, \phantom{aaaaa} {\rm (3D)}
\label{ee_3d}
\end{eqnarray}
and clearly the first is lower for large values of $\lambda$. 
As we see in the next section, this remains true for all coupling strengths for which a polaronic solution exists.

\section{ Numerical Results} 

While the small polaron corresponding to the strong coupling limit can be solved analytically, 
as the coupling strength decreases the electron spreads out over many lattice sites. 
Many different bonds are stretched or compressed to form the accompanying lattice distortion and this 
necessitates numerical calculations to find the lowest energy solution away from strong coupling. 
The important question at weak coupling is to determine if there is a critical coupling strength needed for polaron formation. 
To answer this we started with a calculation at strong coupling and then slowly
lowered the coupling strength in small increments, calculating the low energy 
ion configuration at each step. Each subsequent minimization was seeded with the 
previous slightly higher lambda configuration and we repeated the process for many 
cluster sizes to converge finite size effects. Even in situations where the polaron configuration remains a local
minimum in the energy landscape, if the polaron energy is higher than the electronic energy in the presence of an
undistorted lattice, this signals the presence of a critical coupling strength below which the electron prefers to 
reside in a Bloch wave state surrounded by an undistorted ion lattice.

In 1D, as shown in Fig.~(\ref{fig:1Danalytical}), we found that there was no critical coupling for polaron formation, i.e.
the polaron energy remains lower than the bare tightbinding energy (for $k=0$) for all coupling strengths down to $\lambda = 0 $. 
Note that we used several lattice sizes, and for sizes beyond $30$ or so, finite size effects have disappeared.

\begin{figure}
\begin{center}
\includegraphics[height=2.8in,width=3.7in]{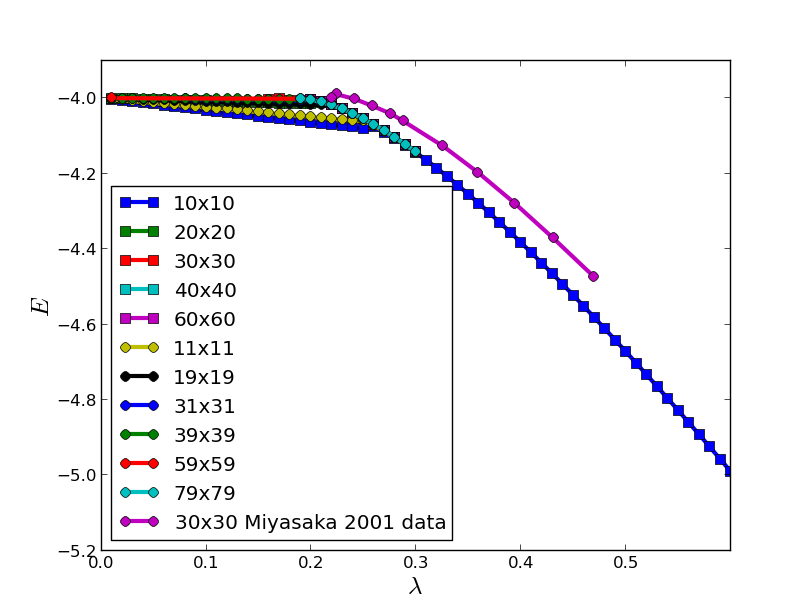}
\caption{ The energy for the 2D BLF-SSH model from numerical calculations (periodic boundary conditions) for a wide range of coupling strengths and 
using both possible starting configurations. Finite size effects have been converged to the thermodynamic limit by using larger and larger clusters. The 
previous results of Miyasaka and Ono\cite{miyasaka01} have been extracted from their 2001 paper to show that our energies are somewhat lower. }
\label{fig:ssh2D}
\end{center}
\end{figure}
\begin{figure}
\begin{center}
\includegraphics[height=2.8in,width=3in]{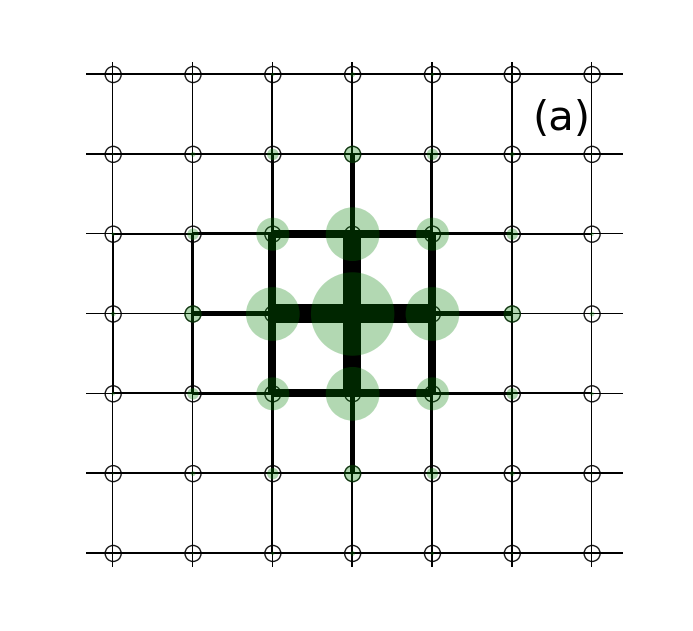}
\includegraphics[height=2.8in,width=3in]{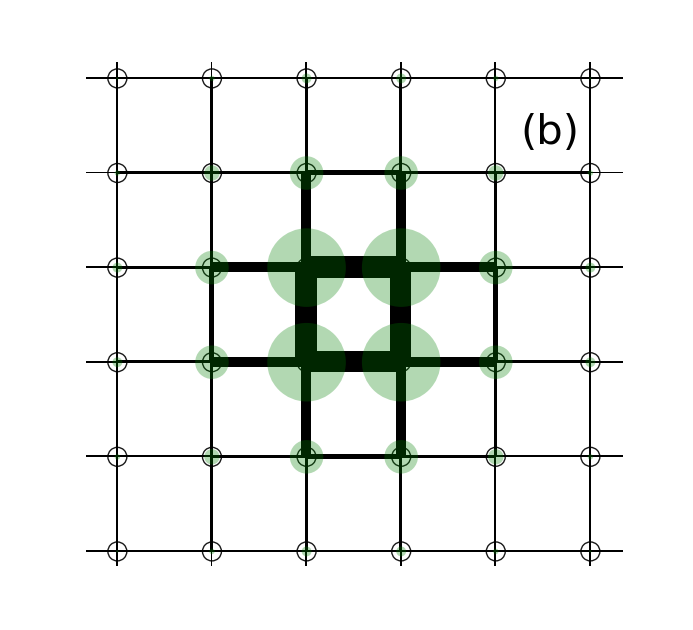}
\caption{ The electron and ion distortion configuration at $\lambda = 0.4$ for the two configurations discussed in the text. In (a) and (b) we show
the `star' and `square' configurations, respectively. The energies of both these configurations are degenerate. The width of the black rectangles connecting 
the ion sites is proportional to the amount that a bond is compressed. Either of these configurations differs qualitatively from the quasi-one-dimensional
configuration found in Ref. (\onlinecite{miyasaka01}).}
\label{fig:config2D}
\end{center}
\end{figure}

The two dimensional results are shown in Fig.~(\ref{fig:ssh2D}) for a variety of lattice sizes, as indicated. 
Here there is no question that a critical point occurs, at $\lambda_{\rm c} = 0.18$. Note that the energies for the
two configurations (`star' and `square') remain degenerate down to this critical coupling strength.
This critical point therefore
occurs for both polaron configurations discussed above (and in the Appendix). A snapshot of the electron and ion configuration 
for a modest coupling strength ($\lambda = 0.4$) is shown in Fig.~(\ref{fig:config2D})(a)and (b) for 
the `star' and `square' configuration, respectively. Even though these configurations have the same energy, they represent
quite different electron and ion distortion patterns. Note, however, that both these solutions have the full symmetry of the 
underlying lattice, and differ considerably from the solution reported in Ref.~(\onlinecite{miyasaka01}).

There are significant finite size effects but we were able to model large enough clusters to eliminate these. 
When the cluster size is small a notable distortion remains, even in the weak coupling limit. This can be understood
by using the following energy, written for the case where all the bonds in the cluster are slightly stretched by the same amount
(written as $y$ in dimensionless units): 
\begin{equation}
 E = 4( -t + \lambda W y) + N\lambda W y^2 
\end{equation}
Minimization with respect to $y$ yields $y = -2/N$, and results in a minimum energy
\begin{equation}
 E = \frac{-4\lambda W }{N} -4t.
\end{equation}
Of course in the thermodynamic limit, $ N \rightarrow \infty$ so the distortion goes to zero, as
is apparent in Fig.~(\ref{fig:ssh2D}).

\begin{figure}
\begin{center}
\includegraphics[height=2.8in,width=3.7in]{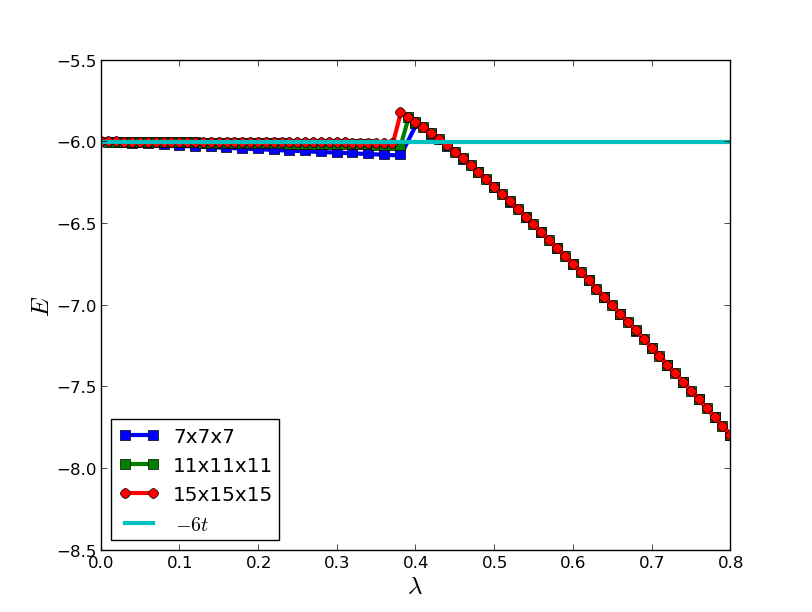}
\caption{ The energy for the 3D BLF-SSH model from numerical calculations (periodic boundary conditions) for a wide range of coupling 
strengths. }
\label{fig:ssh3D}
\end{center}
\end{figure}

In three dimensions finite size effects are not so pronounced. We show results for the ground state energy 
in Fig.~(\ref{fig:ssh3D}) for several lattice sizes. We extract
a critical coupling strength of $\lambda_{\rm c} = 0.44$. In some sense the polaron in 3D remains small over all coupling strengths
beyond this critical value, as even for a $ 7 \times 7 \times 7 $  cluster, finite size effects are almost negligible, and this lattice size is
sufficiently large to contain all the ionic distortions present.

\section{Summary} 
We have presented results for the one, two, and three dimensional BLF-SSH model in the
adiabatic limit. The nature of these solutions is not so different from what they were for the Holstein
model.\cite{kabanov93} In one dimension the solution is always polaronic; no matter how weak the electron-ion
coupling, a lattice distortion always accompanies the electron, even though it exists in a Bloch state.
In two and three dimensions there exists a critical coupling strength, $\lambda_c$, below which there
is no longer an ionic distortion, i.e. the electron is completely independent of the ions. This is exactly what
happened with the Holstein model,\cite{kabanov93} and as in that case, we fully expect the quantum solution
to {\it not} display this same behaviour, i.e. we expect that the quantum solution does not have a critical point. Instead,
we expect that a crossover will occur. It remains to be seen how sharp this crossover will be as the characteristic phonon
energy becomes much smaller than the characteristic electron energy. For reference, in the Holstein model it remained
{\it very} sharp.\cite{li11}

The function optimization required here, with many parameters, is quite a difficult problem, more
difficult than was the case with the Holstein problem, which we have also solved. We have taken a number of 
important steps to obtain the true global minimum solutions, and have 
found lower energy configurations than previous works.\cite{miyasaka01} Hopefully these precise solutions
will help to inform further quantum mechanical studies of the BLF-SSH model and many-electron 
calculations in the adiabatic limit. The small polaron formation in 3D is particularly interesting 
since many studies in semiconductors have used large polarons and Fr\"ohlich-like models. However, for atomic 
semiconductors such as silicon, the Einstein oscillators or optical phonons used in most of the Fr\"ohlich-like models simply do not exist. 
Further research and full quantum mechanical solutions are necessary to resolve the role of small vs large polarons,
to see if the `sharpness' of the crossover from weak to strong coupling present in the Holstein model\cite{li11}
remains for the BLF-SSH model. In any event, we fully expect the strong coupling solutions obtained here to faithfully
reflect the fully quantum mechanical solutions in strong coupling, via coherent states.

\begin{acknowledgments}

This work was supported in part by the Natural Sciences and Engineering
Research Council of Canada (NSERC),by the Alberta iCiNano program, and by Alberta Ingenuity.
\end{acknowledgments}

\appendix
\section{ Strong coupling limit for the adiabatic approximation in two and three dimensions}

In two dimensions the possible configurations are as shown in Fig.~(\ref{fig:config2D}). For the first (star), the electronic
wave function is given as a linear combination of the central site and the symmetric combination of the four surrounding
sites,
\be
|\phi_{\rm star} \rangle  =  a_0  c_{00}^\dagger|0\rangle + a_1 {1 \over 2} (c_{10}^\dagger + c_{-10}^\dagger + c_{01}^\dagger + c_{0-1}^\dagger) | 0\rangle.
\label{basis_star}
\ee
Evaluating the relevant matrix elements results in a $2\times 2$ eigenvalue problem,
   \begin{equation}
   \left (
   \begin{array}{cc}
   0 & -2(t - \lambda W u_0) \\
   -2  (t - \lambda W u_0) & 0
   \end{array}
   \right )
   \left (
   \begin{array}{cc}
   a_0 \\
   a_1
   \end{array} 
   \right )
   = \epsilon_{\rm el}
   \left (
   \begin{array}{cc}
   a_0 \\
   a_1
   \end{array} 
   \right ),
   \end{equation}
where, due to the symmetry of the problem, the dimensionless stretches on the left and right of the centre will be equal to the
stretches above and below the centre (denoted here by $u_0$).  The eigenvalues are readily determined, and when combined
with the ionic energy, results in $u_0 = -1/2$, so that the total energy is
\be
E_{\rm star} = -{1 \over 2}\lambda W - 2t. \phantom{aaaa} {\rm (2D)}
\label{ee_star}
\ee
This constitutes the strong coupling solution for the star configuration in two dimensions. 
We now turn to the `competing' symmetry, the so-called `square' configuration, as depicted also in Fig.~(\ref{fig:config2D}).
Here symmetry dictates that there is only one electronic wave function, a linear combination of the electron located at each of the four
corners:
\be
|\phi_{\rm star} \rangle  =  {1 \over 2} (c_{00}^\dagger + c_{10}^\dagger + c_{11}^\dagger + c_{01}^\dagger) | 0\rangle.
\label{basis_square}
\ee
We for the hopping which is modulated by an longitudinal ionic distortion which we will denote by $v_0$ and which is the same in all directions. 
We find an electronic energy $\epsilon_{\rm el} = -2(t - \lambda W v_0)$. When combined with the ion energy, minimization leads to $v_0 = -1/2$
and
\be
E_{\rm square} = -{1 \over 2}\lambda W - 2t. \phantom{aaaa} {\rm (2D)}
\label{ee_square}
\ee
This is in complete agreement with the energy of the star configuration. One can proceed further with bond distortions in either case
extending further from the central region, but neither configuration can be solved in closed form. Remarkably, numerical diagonalization
leads to results that are numerically indistinguishable nonetheless. 

A similar exercise for the honeycomb lattice, however, results in a strong coupling solution of $E_{\rm  star} = -\frac{1}{2} \lambda W - \sqrt{3}t$
and $E_{\rm hex } = -\frac{1}{3} \lambda W - 2t$, where the `star' configuration, with one centrally located electron amplitude (like the
`star' configuration noted above) has a lower energy than the `hexagonal' configuration, which has six sites occupied by the electron with
equal amplitude. Once again, these strong coupling solutions can be further developed as a power series in $1/\lambda$ by including more sites.

Finally, the same exercise can be performed in three dimensions, for a cubic system; the two competing configurations are the `star'
configuration with one central electron amplitude surrounded by six nearest neighbour amplitudes, and the `cube' configuration, consisting
of the eight sites constituting the corners of the cube having an equal amplitude for the electron (so, as in the two dimensional `square' and
`hexagonal' configurations, the centre of the polaron is {\it not} a lattice site).
One finds, for three dimensions in the strong coupling limit,
\begin{eqnarray}
E_{\rm star}& = &-\frac{1}{2}\lambda W - \sqrt{6}t  \phantom{aaaa} {\rm (3D)} \nonumber \\
E_{\rm cube}& = &-\frac{3}{8}\lambda W - 3t, \phantom{aaaaa} {\rm (3D)}
\label{ee_3d_app}
\end{eqnarray}
so, in the strong coupling limit the `star' configuration has lower energy than the `cube' configuration. As described in the text, by solving the
problem numerically, we have found that 
this remains true over all coupling strengths for which a polaronic configuration with accompanying lattice distortions is the ground state.


\begin{thebibliography}{99}

\bibitem{barisic70} S. Bari\u si\' c, J. Labb\' e, and J. Friedel, Phys. Rev. Lett.\textbf{25}, 919 (1970); S. Bari\u si\' c, Phys. Rev. B\textbf{5}, 932 (1972), S. Bari\u si\' c, Phys. Rev. B\textbf{5}, 941 (1972).

\bibitem{su79} W.P. Su, J.R. Schrieffer, and A.J. Heeger, Phys. Rev. Lett.\textbf{42}, 1698 (1979).

\bibitem{su80} W.P. Su, J.R. Schrieffer, and A.J. Heeger, Phys. Rev. B\textbf{22}, 2099 (1980).

\bibitem{csgmodel} This modelling with optical phonons seems to have started with M. Capone, W. Stephan and M. Grilli, Phys. Rev. B\textbf{56}, 4484 (1997), and more recently with D.J.J. Marchand, G. De Filippis, V. Cataudella, M. Berciu, N. Nagaosa, N.V. Prokof'ev, A.S. Mishchenko, and P.C.E. Stamp, Phys. Rev. Lett. {\bf 105}, 266605 (2010) and V. Cataudella, G. De Filippis, and C.A. Perroni, Phys. Rev. B{\bf 83}, 165203, (2011).

\bibitem{solar} Pedro H. de Oliveira Neto, Wiliam F. da Cunha, Luiz F. Roncaratti, Ricardo Gargano, Geraldo M. e Silva, Chemical Physics Letters \textbf{493} 283–287 (2010).

\bibitem{biophysics} Guiqing Zhang, Peng Cui, Jian Wu, Chengbu Liu, Physica B \textbf{404}  1485–1489 (2009)

\bibitem{holstein59} T. Holstein, Ann. Phys. (New York) \textbf{8}, 325 (1959).

\bibitem{hirsch82} J.E. Hirsch and E. Fradkin, Phys. Rev. Lett.\textbf{49}, 402 (1982); E. Fradkin and J.E. Hirsch, Phys. Rev. B\textbf{27}, 1680 (1983).

\bibitem{peeters85} F.M. Peeters and J.T. Devreese, Phys. Rev. B{\bf 32}, 3515 (1985); W. B. da Costa and F.M. Peeters,
J. Phys. Condens. Matter {\bf 8}, 2173 (1996); G.A. Farias, W.B. da Costa and F.M. Peeters, Phys. Rev. B{\bf 54}, 12835 (1996).

\bibitem{li11} Zhou Li, Carl Chandler, and Frank Marsiglio, Phys. Rev. B \textbf{83}, 045104 (2011). Note that this work
disagrees qualitatively with M. Zoli, Physica C\textbf{384}, 274 (2003), and references therein.

\bibitem{miyasaka01} Norio Miyasaka, and Yoshiyuki Ono, J. Phys. Soc. Jpn. 70 pp. 2968-2976 (2001).

\bibitem{kabanov93} V. V. Kabanov and O. Yu. Mashtakov, Phys. Rev. B \textbf{47} 6060 (1993).

\bibitem{bonca99} J. Bonc\'a, S.A. Trugman and I. Batist\'ic, Phys. Rev. B{\bf 60}, 1633 (1999). 
See also H. Fehske and S.A. Trugman, in Polarons in Advanced
Materials edited by A. S. Alexandrov, Springer Series in Material Sciences
\textbf{103} pp. 393-461, Springer Verlag, Dordrecht (2007).

\bibitem{li10} Zhou Li, D. Baillie, C. Blois, and F. Marsiglio, Phys. Rev. B\textbf{81}, 115114 (2010).

\bibitem{berciu13} See C.P.J. Adolphs and M. Berciu, Europhysics Letters {\bf 102}, 47003 (2013), for issues
concerning the validity of a linear expansion.

\bibitem{Conwell00}Conwell E M and Rakhmanova S V, 2000 Proc. Natl. Acad. Sci. USA 97 4556.

\bibitem{DEbook}Differential Evolution: A Practical Approach to Global Optimization, By Kenneth Price, Rainer M. Storn, and
Jouni A. Lampinen.


\end{thebibliography}
\end{document}